\documentstyle[11pt,a4,cite,epsf]{article}

\def\bbox#1{\mbox{\boldmath$#1$}}

\def\corresponds{{\lower.2ex\hbox{=}}{\rm\kern-.75em^\triangle}}
\def\succsim{\succ\kern-.9em_\sim\kern.3em}
\def\precsim{\prec\kern-1em_\sim\kern.3em}
\def\slantfrac#1#2{\kern1em^{#1}\kern-.3em/\kern-.1em_{#2}}
\def\lfrac#1#2{{}^{#1\!}\kern-.0em/_{#2}}

\def\buildrel#1\under#2{\mathrel{\mathop{\kern0pt #2}\limits_{#1}}}

\begin{document}

\bibliographystyle{myprsty}

\begin{center}
{\it This paper is dedicated to Professor Theodor W. H\"{a}nsch on the
occasion of his birthday.}\\
\end{center}
\vspace{0.5cm}
\begin{center}
{\Large \sf Two--Loop Self--Energy Corrections}
{\Large \sf to the Fine--Structure}
\end{center}
\vspace{0.2cm}
\begin{center}
Ulrich D. Jentschura$^{1),2)}$ and Krzysztof Pachucki$^{3)}$
\end{center}
\vspace{0.2cm}
\begin{center}
$^{1)}$ Institut f\"{u}r Theoretische Physik,\\
Technische Universit\"{a}t Dresden, 01062 Dresden, Germany\\[2ex]
$^{2)}$ National Institute of Standards and Technology,\\
Mail Stop 8401, Gaithersburg, Maryland 20899-8401, USA\\[2ex]
$^{3)}$ Institute of Theoretical Physics,
University of Warsaw, \\
ul.~Ho\.{z}a 69, 00-681 Warsaw, Poland\\[2ex]
{\bf Email:} ulj@nist.gov, krp@fuw.edu.pl
\end{center}
\vspace{0.3cm}
\begin{center}
\begin{minipage}{11.8cm}
{\underline{Abstract}}
We investigate
two-loop higher-order binding corrections to the fine structure,
which contribute to the spin-dependent part of the Lamb shift.
Our calculation focuses on the so-called ``two-loop self-energy''
involving two virtual closed photon loops. For bound states,
this correction has proven to be notoriously difficult to evaluate.
The calculation of the binding corrections
to the bound-state two-loop self-energy
is simplified by a separate treatment
of hard and soft virtual photons. The two photon-energy
scales are matched at the
end of the calculation. We explain the significance of the mathematical
methods employed in the calculation in a more general context, and present
results for the fine-structure difference of the two-loop self-energy
through the order of $\alpha^8$.
\end{minipage}
\end{center}
\vspace{1.3cm}

calculations and mathematical techniques\\
in atomic and molecular physics,\\
relativistic and quantum electrodynamic\\
effects in atoms and molecules,\\
fine and hyperfine structure. \\

\newpage

%
%
\section{Introduction}
\label{Introduction}

Ultra-precise measurements in atomic systems represent today
of the most stringent available tests of fundamental quantum 
theories and a means for the determination of fundamental physical
constants with unprecedented accuracy~\cite{NiEtAl2000}.
The theoretical description of the bound states at a level of 
accuracy which matches the current experimental 
precision, which has reached $1.8$ parts in $10^{14}$ and
whose accuracy is to be improved in the near 
future~\cite{HaPr2001}, demands a thorough
understanding of the bound state including -- among other
effects -- the relativistic,
one-loop, two-loop and higher-order
radiative, recoil, radiative-recoil, and nuclear-size 
corrections~\cite{SaYe1990,MoPlSo1998}.

\begin{figure}[htb]
\begin{center}
\begin{minipage}{10cm}
\centerline{\mbox{\epsfysize=7.7cm\epsffile{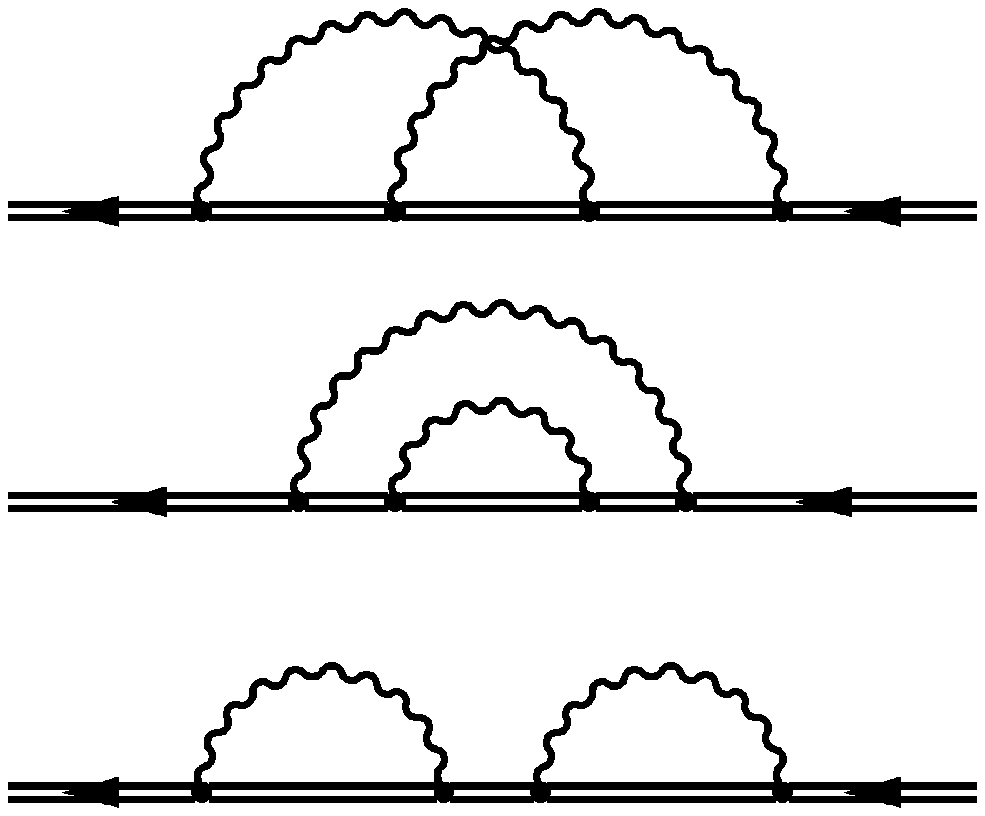}}}
\caption{\label{fig1} Feynman diagrams representing the
two-photon electron self-energy.
The double line denotes the bound electron propagator.
The arrow of time is from right
to left.}
\end{minipage}
\end{center}
\end{figure}

We focus here on radiative corrections, which can be described 
-- for atomic systems with low nuclear charge number -- by a nonanalytic
expansion in powers of the three parameters (i) $\alpha$ (the fine-structure
constant), (ii) the product $Z\alpha$ ($Z$ is the nuclear charge
number), and (iii) the logarithm $\ln[(Z\alpha)^{-2}]$.
The expansion in powers of $\alpha$, which is the perturbation
theory parameter in quantum electrodynamics (QED), corresponds to the 
number of loops in the diagrams. The bound-state effects are taken
into account by the expansions in the two latter parameters.
Higher-order terms in the
expansions in powers of $Z\alpha$ and $\ln[(Z\alpha)^{-2}]$
are referred to as the ``binding corrections''.
One of the historically most problematic sets of Feynman diagrams
in the treatment of the Lamb shift for atomic systems 
has been the radiative correction due to two closed
virtual-photon loops shown in fig.~\ref{fig1}.

%
%

Let us recall at this point that even the evaluation
of higher-order binding corrections to the one-loop self-energy,
which {\em a priori} should represent a less involved calculational
challenge, has represented a problem for analytic evaluations
for over three decades~\cite{ErYe1965ab,Er1971,Sa1981,Pa1993,JeMoSo1999}.
The energy shifts of the bound states due to the radiative 
corrections are conveniently expressed by
expansion coefficients corresponding
to the powers of $Z\alpha$ and $\ln[(Z\alpha)^{-2}]$;
the naming convention is that the power of $Z\alpha$ and 
the power of the logarithm are indicated as indices to the 
analytic coefficients~[see also eq.~(\ref{DefESE})] below.
Because the expansion in both the one-loop and the two-loop
case starts with the fourth power of $Z\alpha$, the non-vanishing
coefficients carry indices $A_{kl}$ and $B_{kl}$ for the 
one- and two-loop cases, respectively 
(with $k\geq 4$ -- see~\cite{SaYe1990} for a comprehensive review).

Logarithmic corrections with $l \geq 1$ can sometimes be inferred
separately in a much simplified approach, e.g.~by considering infrared
divergent contributions to electron form factors. By contrast,
the higher-order non-logarithmic
coefficients represent a considerable calculational challenge.
Realistically, i.e.~with the help of current computer
algebra systems~\cite{Wo1988,disclaimer}, one can hope to 
evaluate non-logarithmic coefficients of sixth order
in $Z\alpha$. Complete results for the one-loop higher-order
correction $A_{60}$ for S and P states have only been
available recently~\cite{Pa1993,JePa1996,JeSoMo1997}.
Calculational difficulties have by now precluded
a successful evaluation of the corresponding coefficient
$B_{60}$ for the two-loop effect. Ground-work for the
evaluation of $B_{60}$ was laid in~\cite{Pa2001}.
Here, we are concerned with the evaluation of the fine-structure
differences of the logarithmic and non-logarithmic coefficients
$B_{6L}$ (where $L = 0,1,2$), i.e.~with the 
$n{\mathrm P}_{3/2}$--$n{\mathrm P}_{1/2}$ difference of these
coefficients.

Using natural Gaussian units 
($\hbar = c = \epsilon_0 = 1$), as it is customary for the current type of
calculation, we write the two-photon self-energy in the 
$Z\alpha$-expansion for P states in terms of $B$-coefficients as
\begin{eqnarray}
\label{DefESE}
\lefteqn{\Delta E_{\mathrm{SE}} = \left(\frac{\alpha}{\pi}\right)^2 \,
(Z\alpha)^4 \, \frac{m}{n^3} \, \bigg[ B_{40} } \nonumber\\[1ex]
& & \quad + (Z\alpha)^2 \,
\left[ B_{62} \, \ln^2(Z\alpha)^{-2} 
+ B_{61} \, \ln(Z\alpha)^{-2} + B_{60} \right] + {\mathcal R} \bigg]\,,
\end{eqnarray}
where the remainder ${\mathcal R}$ is of order 
${\mathcal O}(Z\alpha)^3$.
Relevant Feynman diagrams are shown in fig.~\ref{fig1}.

Here, $m$ denotes the electron mass (we write
eq.~(\ref{DefESE}) in the non-recoil limit, i.e.~for an infinite
nuclear mass). The double logarithmic $B_{62}$-coefficient 
is spin-independent, so that
we have $\Delta_{\mathrm{fs}} B_{62} = 0$.
In this paper, we evaluate the fine-structure
differences
\begin{eqnarray}
\Delta_{\mathrm{fs}} B_{61} &=& 
  B_{61}(n {\mathrm P}_{3/2}) - B_{61}(n {\mathrm P}_{1/2})\,,
\nonumber\\[1ex]
\Delta_{\mathrm{fs}} B_{60} &=& 
  B_{60}(n {\mathrm P}_{3/2}) - B_{60}(n {\mathrm P}_{1/2})\,. 
\end{eqnarray}
Throughout the paper, we will follow the convention that
$\Delta_{\mathrm{fs}} X \equiv X(n {\mathrm P}_{3/2}) - 
X(n {\mathrm P}_{1/2})$ denotes the ``fine-structure part'' of a
given quantity $X$.
For $\Delta_{\mathrm{fs}} B_{61}$ and $\Delta_{\mathrm{fs}} B_{60}$,
we provide complete results.
It is perhaps worth noting that two-loop self-energy effects for bound states 
have represented a considerable challenge for theoretical evaluations.
Our investigation represents a continuation of previous 
work on the two-loop problem 
(see e.g.~\cite{ApBr1970,Pa1994prl,EiKaSh1995,Pa2001}).
It is probably a triviality to express that
technical difficulties in the calculation and its description
in the following sections of the paper
cannot be avoided. 

For the description of the self-energy radiative effects -- mediated by hard
virtual photons --, we use the modified Dirac hamiltonian
\begin{equation}
\label{HDm}
H_{\mathrm D}^{(m)} = \bbox{\alpha} \cdot 
  \left[\bbox{p} -{\mathrm e} \, F_1(\Delta) \, \bbox{A}\right]
     + \beta\,m + {\mathrm e} \, F_1(\Delta) \, \phi +
       F_2(\Delta) \, \frac{e}{2\,m} \, \left({\mathrm i}\, 
         \bbox{\gamma} \cdot
         \bbox{E} - \beta \, \bbox{\sigma} \cdot \bbox{B} \right)\,,
\end{equation}
which approximately describes an electron subject to an
external scalar potential $\phi \equiv \phi(\bbox{r})$
and an external vector potential $\bbox{A} \equiv \bbox{A}(\bbox{r})$. 
This modified hamiltonian is still local in coordinate space. 
The Dirac matrices in (\ref{HDm}) are to be understood in the
standard (Dirac) representation~\cite{ItZu1980} (in the sequel, we will also
use the non-covariant notation $\beta \equiv \gamma^0$ and 
$\alpha^i \equiv \gamma^0 \gamma^i$). 

The argument $\Delta$ of the electron form factors $F_1$ and $F_2$
in eq.~(\ref{HDm}) 
is to be interpreted as a Laplacian operator acting on all
quantities to the right (but not on the wave function of the 
bound electron in evaluating $H_{\mathrm D}^{(m)} | \psi \rangle$). 
In momentum space, the action of the hamiltonian
$H_{\mathrm D}^{(m)}$ is described by the convolution 
$\left[H_{\mathrm D}^{(m)} \psi\right](\bbox{p}')
= \int {\mathrm d}^3 p/(2\pi)^3 \, 
H_{\mathrm D}^{(m)}(\bbox{p}'-\bbox{p}) \, \psi(\bbox{p})$.
The form factors -- in momentum space -- assume arguments according to the 
replacement $\Delta \to - \bbox{q}^2 \equiv - (\bbox{p}'-\bbox{p})^2$.
In eq.~(\ref{HDm}), radiative corrections are taken into account in the 
sense of an effective theory via the inclusion of the
on-shell form factors $F_1$ and $F_2$. Although the bound electron
is not an on-shell particle, the modified hamiltonian
(\ref{HDm}) can still approximately account for significant
radiative systems with low nuclear charge number $Z$. 
Of course, the hamiltonian (\ref{HDm})
cannot offer a complete description of the bound electron.
Recoil effects cannot be described by a one-particle 
equation {\em in principle}, and vacuum-polarization
effects are not contained in eq.~(\ref{HDm}). However, the effective description
of self-energy radiative corrections mediated by hard virtual photons given by 
eq.~(\ref{HDm}) will turn out to be useful in the context
of the current investigation.
 
Both of the form factors $F_1$ and $F_2$ entering in
eq.~(\ref{HDm}) are infrared divergent, but this divergence
is cut off in a natural way at the atomic 
binding energy scale $(Z\alpha)^2\,m$.
The fact that on-shell form factors can describe radiative corrections
to the fine structure --
mediated by high-energy virtual photons -- has been demonstrated explicitly
in~\cite{Pa1999}. The modified Dirac hamiltonian (\ref{HDm}) and 
the associated modified Dirac equation have been introduced 
-- in the one-loop approximation -- in ch.~7 of~\cite{ItZu1980} 
[see for example eqs.~(7-77) and (7-103) {\em ibid.}].
The low-energy part of the calculation is carried out using 
nonrelativistic approximations in the spirit of the simplified
treatment introduced in the previous one- and two-loop
calculations~\cite{Pa1993,JePa1996,JeSoMo1997,Pa1998,Pa2001}.
This approach was inspired, in part, by various attempts to 
formulate simplified low-energy (``nonrelativistic'') approximations
to quantum electrodynamics (``NRQED''), see 
e.g.~\cite{BB1984,CaLe1986}. Both the
high-energy and the low-energy contributions are matched at the 
separation scale $\epsilon$ whose r\^{o}le in the calculation
is illustrated by the mathematical model example discussed
in app.~\ref{appa}.

In a two-loop calculation, either of the two virtual
photons may have a high or low energy as compared to the 
separation scale $\epsilon$.
{\em A priori}, this necessitates~\cite{Pa2001} a separation 
of the calculation into three different contributions: (i) both
photon energies large, (ii) one photon with a large and one with a 
small energy, and (iii) both photons with small energies. 
For the particular problem at hand (the fine-structure
differences of $B_{61}$ and $B_{60}$), we are in the fortunate position
that effects caused by hard virtual photons (i) are described by the
modified Dirac hamiltonian (\ref{HDm}), whereas the low-energy
part discussed in sec.~\ref{lep} below comprises both  
remaining contributions (ii) and (iii). 

This paper is organized as follows: Two-loop form factors entering in
eq.~(\ref{HDm}) are analyzed in sec.~\ref{twoloopff}. The calculation
is split into two parts: the high-energy part discussed in sec.~\ref{hep}
and the low-energy part, which is treated along ideas 
introduced in~\cite{CaLe1986} in sec.~\ref{lep}. Results and conclusions
are left to sec.~\ref{resconcl}.  

%
%
\section{Two-loop Form Factors}
\label{twoloopff}

In order to analyze the modified Dirac hamiltonian (\ref{HDm}) 
through two-loop order, we first have to investigate certain 
expansion coefficients of the electronic 
$F_1$ and $F_2$ form factors which are
thoroughly discussed in the seminal papers~\cite{Re1972a,Re1972b}. 
For the momentum transfer $q^2$ which is the argument of the 
two functions $F_1 \equiv F_1(q^2)$ and $F_2 \equiv F_2(q^2)$,
we use the convention $q^2 = q_\mu q^\mu = (q^0)^2 - \bbox{q}^2$.
The variable $t$ in~\cite{Re1972a,Re1972b} is given 
as $t = q^2$. When we evaluate radiative corrections to the 
binding Coulomb field which is mediated by space-like virtual
photons, we have $q^2 = - \bbox{q}^2$ because for $q^0 = 0$.
We use the conventions (see eq.~(1.2) in~\cite{Re1972a}):
\begin{equation}
F_1(t) = 1 + \sum_{n=1}^{\infty} \left(\frac{\alpha}{\pi}\right)^n
  F_1^{(2n)}(t) \,,\quad
F_2(t) = \sum_{n=1}^{\infty} \left(\frac{\alpha}{\pi}\right)^n
  F_2^{(2n)}(t)\,.
\end{equation}
One and two-loop effects are denoted by upper 
indices 2 and 4, respectively. This notation is motivated
by the observation that two-loop effects are of forth 
order in the quantum electrodynamic interaction Lagrangian
$- e\, \bar{\psi}\, \gamma^\mu A_\mu \, \psi$ 
(in the Furry picture, which is used for the description of bound states,
the Coulomb interaction is taken out of the interaction Lagrangian). 

\begin{figure}[htb]
\begin{center}
\begin{minipage}{12cm}
\centerline{\mbox{\epsfysize=6.7cm\epsffile{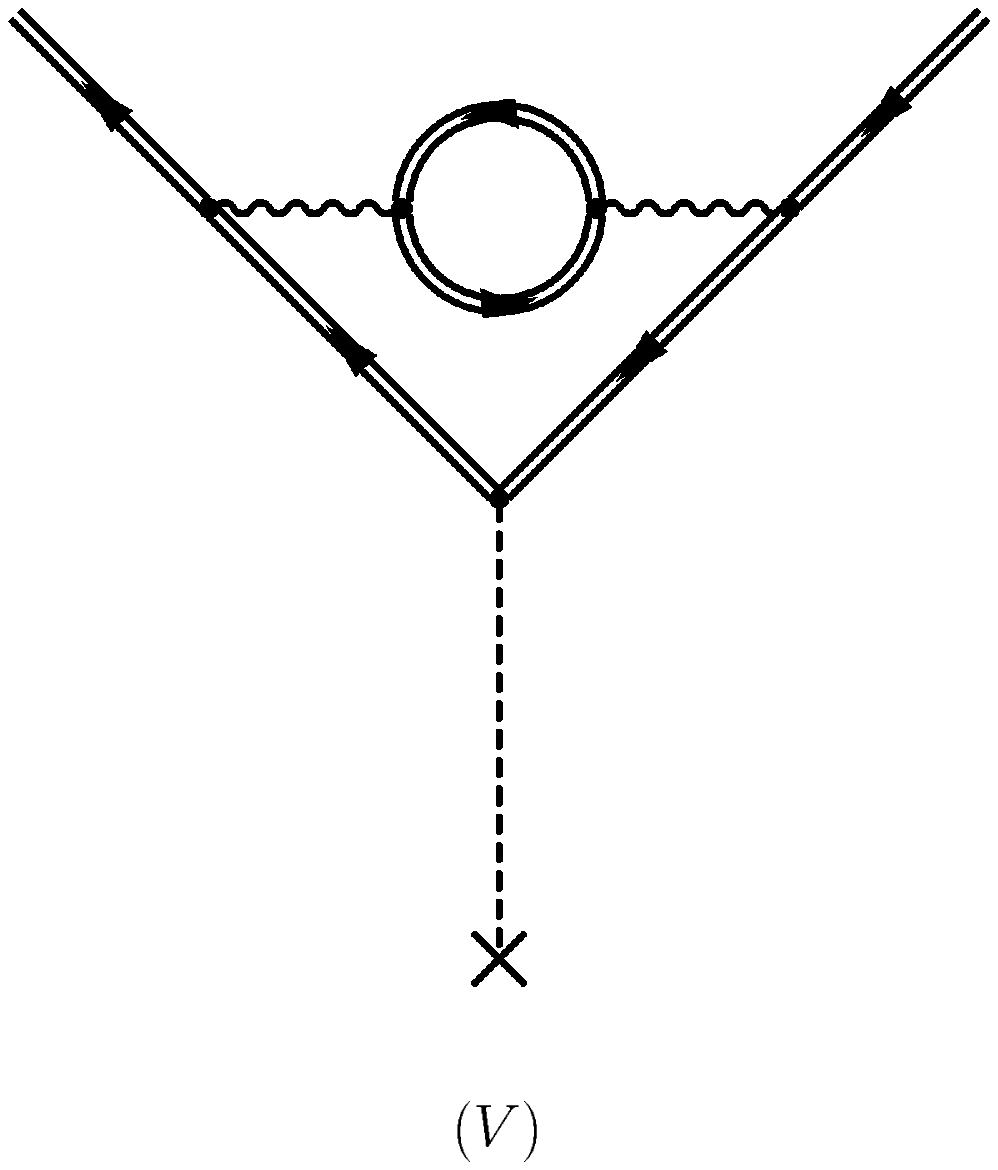}}}
\caption{\label{fig2} Combined self-energy vacuum-polarization diagram
(denoted ``$V$'' in the text).}
\end{minipage}
\end{center}
\end{figure}

There are two different points of view regarding the choice of
diagrams to be included in the two-loop form factors, depending on 
whether the self-energy vacuum polarization diagram~fig.~\ref{fig2}
is included in the calculation or not. We will discuss both cases
and give results with and without the diagram shown in~fig.~\ref{fig2}
taken into account.  

%
%

First, we discuss results obtained for $F_1$ {\em including} the
combined self-energy vacuum polarization diagram. In this case, the
known results for the slope $F_1'(0)$  
and for $F_2(0)$, through two-loop order, read as follows. From
eq.~(1.11) of~\cite{Re1972a}, we have:
\begin{eqnarray}
\lefteqn{m^2 F_1'(0) = \frac{\alpha}{\pi} \, 
\left[- \frac{1}{3} \, \ln\left(\frac{\lambda}{m}\right) -  
  \frac{1}{8}\right] }\nonumber\\[1ex] 
& & \quad + \left(\frac{\alpha}{\pi}\right)^2 \, 
  \left[ - \frac{4819}{5184} - \frac{49}{72} \, \zeta(2) +
    3\,\zeta(2)\,\ln 2 - \frac{3}{4} \, \zeta(3) \right]\,,
\end{eqnarray}
where the forth-order coefficient has the numerical value
\begin{equation}
\label{resultF1prime}
m^2 F_1'^{(4)}(0) = 0.469\,941\,487\,460\,.
\end{equation}
According to eq.~(1.7) in~\cite{Re1972a}, the value of $F_2(0)$,
through two-loop order, reads
\begin{equation}
F_2(0) = \frac{1}{2}\,\frac{\alpha}{\pi} 
+ \left(\frac{\alpha}{\pi}\right)^2 \, 
  \left[ \frac{197}{144} + \frac{1}{2} \, \zeta(2) 
   - 3 \, \zeta(2) \, \ln 2 + \frac{3}{4}\,\zeta(3) \right]\,,
\end{equation}
where the two-loop coefficient has the numerical value
\begin{equation}
\label{resultF2}
F_2^{(4)}(0) = -0.328\,478\,965\,579\,.
\end{equation}
We now turn to the discussion of the slope $F_2'^{(4)}(0)$.
In view of eq.~(1.20) of~\cite{Re1972a} (see also~\cite{YeFrSu1961}),
we have (up to two-loop order)
\begin{equation}
\label{yefrsu}
F_2(t) = \frac{\alpha}{\pi} \, {\mathcal F}_2^{(2)}(t) +
  \left(\frac{\alpha}{\pi}\right)^2 \,
     \left[ \ln \frac{\lambda}{m} \, B(t) \, {\mathcal F}_2^{(2)}(t)
        + {\mathcal F}_2^{(4)}(t) \right]\,,
\end{equation}
where the coefficients ${\mathcal F}$ are by definition infrared safe and
\begin{equation}
\label{specificf2}
{\mathcal F}_2^{(2)}(0) = \frac{1}{2} \,,\quad
B(t) = - \frac{t}{3\,m^2} - \frac{t^2}{20\,m^2} +
  {\mathcal O}(t^3)\,.
\end{equation}
Equations (\ref{yefrsu}) and (\ref{specificf2})
uniquely determine the infrared divergent contribution to $F_2'^{(4)}(0)$.
An analytic expressions for ${\mathcal F}_2^{(4)}(t)$, $t$ spacelike, 
has recently been obtained~\cite{MaRe2001} in terms of
harmonic polylogarithms~\cite{ReVa2000,GeRe2001}. As a byproduct, 
an analytic expression for the the slope ${\mathcal F}_2'^{(4)}(0)$
was found. The result reads
\begin{eqnarray}
\label{resultF2prime}
m^2 \, F_2'^{(4)}(0) &=& 
  - \frac{1}{6} \, \ln\left(\frac{\lambda}{m}\right)
    + m^2 \, {\mathcal F}_2'^{(4)}(0)\,, \nonumber\\[2ex]
m^2 \, {\mathcal F}_2'^{(4)}(0) &=& 
\frac{1751}{2160} 
+ \frac{13}{20} \, \zeta(2)
- \frac{23}{10} \, \zeta(2) \, \ln 2
+ \frac{23}{40} \, \zeta(3)\,.
\end{eqnarray}
A numerical result for ${\mathcal F}_2'^{(4)}(0)$, complementing
the above analytic expression,
can easily be derived in combining eq.~(1.20), eq.~(1.30),
and eq.~(3.2) in~\cite{Re1972a}, as will be explained in 
the sequel. The dispersion relation (1.30) in~\cite{Re1972a} reads,
\begin{equation}
\label{dispersion}
{\mathrm{Re}} \, F_2(t) = - \frac{4 m^2}{t - 4 m^2} \, F_2(0) +
\frac{1}{\pi} \, \frac{t}{t - 4 m^2} \, P \int\limits_{4 m^2}^{\infty}
\frac{{\mathrm d} t'}{t' - t} \, \frac{t' - 4 m^2}{t'} \,
{\mathrm{Im}} F_2(t')\,,
\end{equation}
where $P$ denotes the Cauchy principal value.
Equation~(\ref{dispersion}) applies also if we single out the 
two-loop effect and differentiate at zero momentum transfer, 
and we obtain for the slope $F_2'^{(4)}(0)$ the relation
\begin{equation}
\label{F2prime1}
m^2 \, F_2'^{(4)}(0) = \frac{1}{4} \, F_2^{(4)}(0) +
\frac{1}{4 \, \pi} \, P \int\limits_{4 m^2}^{\infty}
{\mathrm d} t' \, \frac{4 m^2 - t'}{t'^2} \,
{\mathrm{Im}} F_2'^{(4)}(t') = 
\frac{1}{4} \, F_2^{(4)}(0) + {\mathcal T}\,,
\end{equation}
where $F_2^{(4)}(0)$ is given in eq.~(\ref{resultF2}).
The second term on the right-hand side, denoted by ${\mathcal T}$,
can be evaluated using the result for ${\mathrm{Im}} \, F_2^{(4)}(x)$
presented in eq.~(3.2) in~\cite{Re1972a}; it reads
\begin{equation}
\label{F2prime2}
{\mathcal T} = - \int\limits_0^1 {\mathrm d} x \frac{(1-x)^3}{x\,(1+x)}\,
  {\mathrm{Im}} \, F_2^{(4)}(x) = 
  - \frac{1}{6} \, \ln\left(\frac{\lambda}{m}\right) +
     0.030\,740\,507\,833(1) \,.
\end{equation}
Here, the last error is due to numerical integration,
and use is made of the natural variable~\cite{Re1972a}
\begin{equation}
x = \frac{1 - \sqrt{1 - 4 m^2/t}}{1 + \sqrt{1 - 4 m^2/t}}\,.
\end{equation}
In combining the result of eq.~(\ref{resultF2}) with 
eqs.~(\ref{F2prime1}) and (\ref{F2prime2}), the result
$m^2 \, {\mathcal F}_2'^{(4)}(0) = - 0.051\,379\,233\,561(1)$
is obtained which is in agreement with (\ref{resultF2prime}).

Now we will provide results for the form factors
obtained {\em excluding} the self-energy vacuum-polarization
graph $V$ shown in fig.~\ref{fig2}. These results refer to the 
pure two-photon self-energy diagrams shown in fig.~\ref{fig1}.
The two-loop self-energy diagrams independently form a gauge-invariant set.
They represent a historically problematic correction, and are the main
subject of our investigation.
The combined self-energy vacuum-polarization diagram,
according to eqs. (1.9) and (1.10) in~\cite{Re1972b} --
taking into account the subtracted dispersion relation (1.30)
of~\cite{Re1972a} -- leads to the following corrections:
\begin{eqnarray}
F_1'^{(4),V}(0) = 
  -\frac{1099}{1296} + \frac{77}{144} \, \zeta(2) =
  0.031\,588\,972\,474\,,
\nonumber\\[3ex]
F_2^{(4),V}(0) =
  \frac{119}{36} - 2 \, \zeta(2) =
  0.015\,687\,421\,859\,,
\nonumber\\[3ex]
F_2'^{(4),V}(0) =
  \frac{311}{216} - \frac{7}{8} \, \zeta(2) =
  0.000\,497\,506\,323\,.
\end{eqnarray}
For the pure self-energy graphs, which we would
like to denote by the symbol $S$, we therefore obtain the 
following results,
\begin{eqnarray}
\label{seF1p0}
m^2 \, F_1'^{(4),S}(0) &=& 
  - \frac{47}{576} - \frac{175}{144}\,\zeta(2)
  + 3\,\zeta(2) \, \ln 2 - \frac{3}{4} \, \zeta(3) =
  0.438\,352\,514\,986\,,\\[3ex]
\label{seF20}
F_2^{(4),S}(0) &=& 
  - \frac{31}{16} + \frac{5}{2}\,\zeta(2)
  - 3\,\zeta(2) \, \ln 2 + \frac{3}{4} \, \zeta(3) =
  -0.344\,166\,387\,438\,,\\[3ex]
\label{seF2p0}
m^2 \, F_2'^{(4),S}(0)  &=&
  - \frac{1}{6} \, \ln\left(\frac{\lambda}{m}\right)
    - \frac{151}{240} + \frac{61}{40}\,\zeta(2)
    - \frac{23}{10} \, \zeta(2) \, \ln 2 
    + \frac{23}{40} \, \zeta(3)
\nonumber\\[1ex]
&=&
  - \frac{1}{6} \, \ln\left(\frac{\lambda}{m}\right)
    - 0.051\,876\,739\,885 
\equiv
  - \frac{1}{6} \, \ln\left(\frac{\lambda}{m}\right)
    + {\mathcal F}_2'^{(4),S}(0)\,.
\end{eqnarray}
where the latter equality defines ${\mathcal F}_2'^{(4),S}(0)$ in 
analogy with eqs.~(\ref{yefrsu}) and (\ref{resultF2prime}).

%
%
\section{High--Energy Part}
\label{hep}

Based on the modified Dirac hamiltonian (\ref{HDm}),
corrections to the energy of the bound Dirac particle
can be inferred.
We will refer to the energy corrections 
attributable to the $F_1$ and $F_2$ form factors
as $E_1$ and $E_2$, respectively. For $E_1$, we have
\begin{equation}
E_1 = \left< \left[ F_1(-\bbox{q}^2) - 1 \right] \,
  e\,\phi \right>_{\mathrm{fs}} \,,
\end{equation}
where the index fs refers to the fine-structure terms, i.e.~to the 
result obtained by subtracting the value of the matrix element for
a $n{\mathrm P}_{3/2}$ state from the value of the same matrix 
element evaluated on a $n{\mathrm P}_{1/2}$ state.  
A matrix element $\left< A \right>_{\mathrm{fs}}$ of a
given operator $A$ is evaluated as
\[
\left< A \right>_{\mathrm{fs}} \equiv
\left< \psi_{{\mathrm{nP}_{3/2}}}^{+} \bigg| \, A \, \bigg| 
\psi_{{\mathrm{nP}_{3/2}}} \right> -
\left< \psi_{{\mathrm{nP}_{1/2}}}^{+} \bigg| \, A \, \bigg|
\psi_{{\mathrm{nP}_{1/2}}} \right>\,,
\]
where $\psi^{+}$ denotes the hermitian conjugate of the 
Dirac wave function $\psi$ (not the 
Dirac adjoint $\bar{\psi} = \psi^{+} \gamma^0$). 
The Dirac wave functions $\psi$ are expanded in powers of $(Z\alpha)$ up to the
order relevant for the current investigation.
This expansion avoids potential
problems associated with the logarithmic divergence of the Dirac
wave function at the origin. 

For $E_1$, up to the order of $(Z \alpha)^6$, we have 
\begin{equation}
E_1 = 4 \pi Z \alpha \, F_1'^{(4)}(0) \, 
  \left< \delta^{(3)}(\bbox{r}) \right>_{\mathrm{fs}}\,.
\end{equation}
For P states, the nonrelativistic 
(Schr\"{o}dinger) wave function -- the leading term in the 
$Z\alpha$-expansion of the Dirac wave function -- 
vanishes at $\bbox{r} = 0$, but the first relativistic
correction gives a finite contribution, resulting in
\begin{equation}
\langle \delta^{(3)}(\bbox{r}) \rangle_{\mathrm{fs}} = 
  - \frac{n^2 - 1}{4\,n^5} \, (Z\alpha)^5 \, m^3\,.
\end{equation}
This leads -- again up to the order of $(Z \alpha)^6$ --
to the following result for $E_1$,
\begin{equation}
\label{E1}
E_1 = \left(\frac{\alpha}{\pi}\right)^2 \,
  \frac{(Z\alpha)^6}{n^3} \, \left[ - F_1'^{(4)}(0) \,
     \frac{n^2 - 1}{n^2} \right] \, m^3\,.
\end{equation}
Observe that the derivative of the $F_1$ form factor has a physical
dimension of $1/m^2$ in natural units, giving the correct
physical dimension for $E_1$. 
The correction due to $F_2$ in (\ref{HDm}) reads,
\begin{equation}
\label{E2}
E_2 = \langle F_2(-\bbox{q}^2) \,
  \frac{e}{2 m} \, \mathrm{i}\, \bbox{\gamma} \cdot \bbox{E} 
     \rangle_{\mathrm{fs}}\,.
\end{equation}
A particle in an external binding Coulomb
field feels an electric field 
$\bbox{E} = {\mathrm i}\,(Z e) \, \bbox{q}/\bbox{q}^2$ 
-- in momentum space -- or 
$\bbox{E} = - (Z e) \, \bbox{r}/(4 \pi r^3)$
in coordinate space. 
Vacuum polarization corrections to 
$\bbox{E} = - (Z e) \, \bbox{r}/(4 \pi r^3)$
lead to higher-order effects. The correction
$E_2$ splits up in a natural way into two
contributions $E_{2a}$ and $E_{2b}$
which are associated with $F_2(0)$ and the slope $F'_2(0)$,
respectively. $E_{2a}$ reads
\begin{equation}
E_{2a} = \frac{Z \alpha}{2 m} \, F_2^{(4)}(0) \, 
\left< - \mathrm{i} \, 
\frac{\bbox{\gamma} \cdot \bbox{r}}{r^3} \right>_{\mathrm{fs}}\,.
\end{equation}
The evaluation of the matrix element leads to
\begin{equation}
\left< - \mathrm{i} \, 
\frac{\bbox{\gamma} \cdot \bbox{r}}{r^3} \right>_{\mathrm{fs}}
  = \left\{ \frac{(Z\alpha)^3}{n^3} + 
    \left[ \frac{487}{360} + \frac{5}{4 n} - \frac{23}{10 n^2}\right]
       \frac{(Z\alpha)^5}{n^3} \right\} \, m^2\,.
\end{equation}
For the purpose of the current investigation,
the $(Z\alpha)^6$-component of $E_{2a}$ is selected only:
\begin{equation}
\label{E2a}
E_{2a} = \left(\frac{\alpha}{\pi}\right)^2 \,
  \frac{(Z\alpha)^6}{n^3} \, \left[ F_2^{(4)}(0) \,
     \left( \frac{487}{720} + \frac{5}{8 n} - 
        \frac{23}{20 n^2} \right) \right]\,m\,.
\end{equation}
The matrix element $E_{2b}$ can be expressed as
\begin{equation}
E_{2b} = \frac{4 \pi Z \alpha}{2 m} \, F_2'^{(4)}(0) \, 
\langle \bbox{\gamma} \cdot \bbox{q} \rangle_{\mathrm{fs}}\,.
\end{equation}
A transformation into coordinate space leads to
\begin{equation}
\langle \bbox{\gamma} \cdot \bbox{q} \rangle_{\mathrm{fs}} = 
{\mathrm i} \, \left[ \frac{\partial}{\partial \bbox{x}} 
\left(\psi^{+}(\bbox{x}) \bbox{\gamma} \psi(\bbox{x}) \right) 
  \right]_{x=0,{\mathrm{fs}}}
  =  - \frac{n^2 - 1}{n^5} \, (Z\alpha)^5\,m^4\,.
\end{equation}
As a function of the principal quantum number $n$,
the result for $E_{2b}$ reads:
\begin{equation}
\label{E2bVersion1}
E_{2b} = \left(\frac{\alpha}{\pi}\right)^2 \,
  \frac{(Z\alpha)^6}{n^3} \, \left[ - 2 \, F_2'^{(4)}(0) \,
     \frac{n^2 - 1}{n^2} \right]\,m^3\,.
\end{equation}                
This result involves the infrared divergent slope
of the $F_2$ form factor [see eqs.~(\ref{resultF2prime}) and (\ref{seF2p0})].
We are thus faced with 
the problem of matching the infrared divergence of the slope of 
the $F_2$ form factor, expressed in terms of the fictitious photon mass
$\lambda$, with the usual (energy matching parameter) $\epsilon$
introduced originally in~\cite{Pa1993}.
This can be done in two ways: (i) by matching the infrared divergence 
of the rate of soft bremsstrahlung, calculated with a fictitious photon
mass $\lambda$, to a result of the same calculation, carried out with
an explicit infrared cut-off $\epsilon$ for the photon energy.
This way of calculation is described on pp.~361--362 of~\cite{ItZu1980}.
It leads to the result
\begin{equation}
\label{matching1}
\ln \frac{\lambda}{2 \, \epsilon} = - \frac{5}{6}\,.
\end{equation}
The matching procedure (ii) consists in a comparison of the result
of the application of the formalism considered above, and its
application to the high-energy part of the ground state Lamb shift,
which is in leading order given by the infrared divergence of the 
$F_1$ form factor, and the result obtained by direct calculation
of this high-energy part in a non-covariant formalism with an 
explicit energy cut-off $\epsilon$, as it has been carried out 
in~\cite{Pa1993}. This second matching procedure leads to the following
result -- in agreement with (\ref{matching1}) --,
\begin{equation}
\label{matching2}
\ln \frac{m}{\lambda} - \frac{3}{8} = \ln \frac{m}{2 \epsilon} + 
  \frac{11}{24}\,.
\end{equation}
So, we are led to the replacement
\begin{equation}
\label{replacement}
- \ln \frac{\lambda}{m} \to 
  \ln \frac{m}{2 \epsilon} + \frac{5}{6}  
\end{equation}
A comparison with the results in eqs.~(\ref{resultF2prime}), (\ref{seF2p0}),
and~(\ref{E2b}) reveals that the
logarithmic divergence for the fine-structure difference is given by
a term
\begin{equation}
\label{logterm}
  - \frac{n^2 - 1}{3 n^2} \, \ln \frac{m}{2 \epsilon} \,,
\end{equation}
so that we may anticipate at this stage the result for 
$\Delta_{\mathrm{fs}} B_{\mathrm 61}$,
\begin{equation}
\label{DeltaB61}
\Delta_{\mathrm{fs}} B_{\mathrm 61} = 
  - \frac{n^2 - 1}{3 n^2} \,.
\end{equation}
Based on (\ref{E2bVersion1}) and (\ref{replacement}),
we can express $E_{2b}$ in terms of $\epsilon$ and 
${\mathcal F}_2'^{(4)}(0)$,
\begin{equation}
\label{E2b}
E_{2b} = \left(\frac{\alpha}{\pi}\right)^2 \,
  \frac{(Z\alpha)^6}{n^3} \, \left[ 
     - \frac{1}{3}\,\frac{n^2-1}{n^2}\,\ln\frac{m}{2\epsilon} - 
          \left(\frac{5}{18} + 2\,{\mathcal F}_2'^{(4)}(0)\,m^2 \right) \, 
    \frac{n^2-1}{n^2} \right]\,m\,.
\end{equation}

There is a third correction due to the effect of {\em two} one-loop 
corrections on the electron vertices. Because we are only interested
in the fine structure, we isolate the terms which are proportional
to the spin-orbit coupling, and we obtain
\begin{equation}
\label{defE3}
E_3 = \left< \left[2 F_2(0)\right] \, H_{\mathrm{fs}} \,
  \left( \frac{1}{E - H} \right)' \,
\left[2 F_2(0)\right] \, H_{\mathrm{fs}} \right>_{\mathrm{fs}}\,,
\end{equation}
where
\begin{equation}
\label{Hfs}
H_{\mathrm{fs}} = \frac{Z\alpha}{4 m^2 r^3} \, 
  \bbox{\sigma} \cdot \bbox{L}\,,
\end{equation}
and $1/(E-H)'$ is the nonrelativistic, spin-independent
reduced Schr\"{o}dinger--Coulomb Green function~\cite{SwDr1991a,SwDr1991b}.
The only spin-dependence in (\ref{defE3})
occurs in the coupling $\bbox{\sigma} \cdot
\bbox{L}$, and it can be taken into account by an overall factor,
\begin{equation}
\label{spindependence}
\left< (\bbox{\sigma} \cdot \bbox{L})^2 \right>_{\mathrm{fs}} = -3\,.
\end{equation}
We are therefore led to consider the ``spin-independent
version'' of the matrix element which occurs in eq.~(\ref{defE3}) 
and obtain the following result,
\begin{equation}
\label{matE3}
\left< \frac{Z\alpha}{4 m^2 r^3} \left(\frac{1}{E-H}\right)'
\frac{Z\alpha}{4 m^2 r^3} \right>_{\mathrm{nP}} =
  \left( -\frac{227}{8640} - \frac{1}{96 n} 
     + \frac{1}{80 n^2} \right) \, \frac{(Z \alpha)^6 \, m}{n^3}\,.
\end{equation}
The spin-dependence can be easily restored by considering 
eq.~(\ref{spindependence}). The index ``nP'' in eq.~(\ref{matE3}) means 
that the matrix element is evaluated with the nonrelativistic,
spin-independent (Schr\"{o}dinger) wave function. Alternatively,
on may evaluate with either the ${\mathrm{nP}}_{1/2}$ or the
${\mathrm{nP}}_{3/2}$ Dirac wave function and expand up to the leading order
in $(Z\alpha)$.

The evaluation of (\ref{matE3}) can proceed e.g.~by solving the differential 
equation which defines the correction to the wave function induced
by $H_{\mathrm{fs}}$, and subsequent direct evaluation of the 
resulting standard integrals using computer algebra~\cite{Wo1988,disclaimer}. 
The final result for $E_3$ reads,
\begin{equation}
\label{E3}
E_3 = \left( \frac{\alpha}{\pi}\right)^2 \, \frac{(Z\alpha)^6}{n^3} \,
  \left[ \frac{227}{2880} + \frac{1}{32 n} - \frac{3}{80 n^2} \right]\,m\,.
\end{equation}
This concludes the discussion of the high-energy part. 
The final result for the high-energy part is
\begin{equation}
\label{EH}
E_{\mathrm H} = E_1 + E_{2a} + E_{2b} + E_3\,,
\end{equation}
where $E_1$, $E_{2a}$, $E_{2b}$, $E_3$ are given in 
eqs.~(\ref{E1}), (\ref{E2a}), (\ref{E2b}), (\ref{E3}), respectively.

%
%
\section{Low--Energy Part}
\label{lep}

The low-energy part consists essentially of two contributions.
Both effects, denoted here $E_4$ and $E_5$,
can be obtained by a suitable variation 
of the low-energy part of the {\em one-loop} self energy, by considering the 
spin-dependent effects introduced by a 
{\em further} one-loop electron anomalous 
magnetic moment interaction.
The first of the two terms, $E_4$, is caused by spin-dependent
higher-order effects in the one-loop self-energy, which receive
additional corrections due to the anomalous magnetic moment
of the electron.
The second term, $E_5$, is due to an anomalous magnetic moment correction
to the electron transition current, which can also be seen
as a correction to the radiation field of the electron
due to its anomalous magnetic moment.

The leading-order
low-energy part (see~\cite{Pa1993}) reads
\begin{equation}
\label{NREL}
E_{\mathrm L} = -\frac{2 \alpha}{3 \pi m} \,
\int_0^\epsilon {\mathrm d}\omega \, \omega \,
  \left< \phi \left| \bbox{p} \, \frac{1}{H - (E - \omega)} \,
     \bbox{p} \right| \phi \right>\,.
\end{equation}
In order to isolate the fine-structure effects,
we should now consider corrections to the wave function, to the current,
to the hamiltonian and to the energy of the bound state due to the 
spin-dependent relativistic (spin-orbit) hamiltonian 
\begin{equation}
\label{calH}
{\mathcal H} =
F_2(0) \, \frac{e}{2 m} \, {\mathrm i} \, \bbox{\gamma} \cdot \bbox{E} =
\frac{\alpha (Z \alpha)}{4 \pi m} \, 
\frac{-{\mathrm i} \, \bbox{\gamma} \cdot \bbox{r}}{r^3}\,.
\end{equation}
The above hamiltonian ${\mathcal H}$
is the last term in the modified Dirac hamiltonian [right-hand side
of eq.~(\ref{HDm})],
approximated for a particle bound in a Coulomb field with the
$F_2$ form factor evaluated at zero momentum. The electric
field $\bbox{E}$ in (\ref{calH}) corresponds to the binding
Coulomb interaction. The hamiltonian (\ref{calH}) 
describes the modification of the spin-orbit interaction due to 
the anomalous magnetic moment of the electron.

The nonrelativistic limit of ${\mathcal H}$ is the spin-orbit
coupling $H_{\mathrm{fs}}$ given in eq.~(\ref{Hfs}),
multiplied by a factor $2 F_2^{(2)}(0) = \alpha/\pi$
(the additional factor 2 finds an explanation
in~\cite{BeSa1957}).
The resulting hamiltonian
\begin{equation}
\label{Heff}
H_{\mathrm{eff}} = \frac{\alpha}{\pi}\, H_{\mathrm{fs}} =
\frac{\alpha}{\pi}\, \frac{Z\alpha}{4 m^2 r^3} \,
  \bbox{\sigma} \cdot \bbox{L}\,.
\end{equation}
takes into account magnetic vertex corrections in the 
framework of an effective theory. Denoting the variation of  
the expression (\ref{NREL}) mediated by $H_{\mathrm{eff}}$ with the symbol
$\delta_{\mathrm{eff}}$ -- in the spirit of
the notation introduced in~\cite{Pa2001} --, we obtain the contribution
\begin{equation}
\label{defE4a}
E_{4a} = \delta_{\mathrm{eff}} 
\left\{ - \frac{2 \alpha}{3 \pi m} \,
\int_0^\epsilon {\mathrm d}\omega \, \omega \,
  \left< \phi \left| \, \bbox{p} \, \frac{1}{H - (E - \omega)} \,
     \bbox{p} \, \right| \phi \right> \right\}\,.
\end{equation}
Following the notation introduced in~\cite{JePa1996,JeSoMo1997},
the contribution $E_{4a}$ is the sum of the fine-structure effects
created by the wave-function-correction $F_{\delta\phi}$, 
the first relativistic correction to the energy $F_{\delta E}$,
and the correction due to the relativistic hamiltonian $F_{\delta H}$,
each multiplied by a factor $\alpha/\pi$.
That is to say, the final result for $E_4$ is
\begin{equation}
\label{E4aspec}
E_{4a} = \left(\frac{\alpha}{\pi}\right)^2 \, (Z\alpha)^4 \,
\frac{m}{n^3} \, \left(\Delta_{\mathrm{fs}} F_{\delta\phi} + 
\Delta_{\mathrm{fs}} F_{\delta E} +
\Delta_{\mathrm{fs}} F_{\delta H} \right)\,.
\end{equation}
There is a further correction to the nonrelativistic effective coupling
to the radiation field due to the ``anomalous 
spin-orbit hamiltonian'' (\ref{calH}).
The correction, in the nonrelativistic limit, can be derived by considering
a Foldy--Wouthuysen transformation which by definition diagonalizes  
the hamiltonian~(\ref{calH}) in spinor space and also 
leads to a transformation of the relativistic current operator $\alpha^i$ 
according to
\begin{equation}
\label{FWcorrection}
\alpha^i \to U \, \alpha^i U^{-1} \,, \quad
U = \exp\left(-{\mathrm i} \frac{\beta {\mathcal H}}{2 m} \right) \,.
\end{equation}
Here, $\beta$ and $\alpha^i$ are standard Dirac matrices~\cite{ItZu1980},
$i$ is a spatial index, and ${\mathcal H}$ is given in (\ref{calH}).
The calculation is carried
out along ideas introduced in~\cite{JePa1996} and leads to the result
\begin{equation}
\delta \bbox{j}_{4b} = \frac{\alpha}{\pi} \, \frac{Z\alpha}{2 m r^3} \,\,
\bbox{\sigma} \times \bbox{r}\,,
\end{equation}
as a relativistic correction to the electron current
which is simply $\alpha^i$ in the relativistic
formalism and  $p^i/m$ in the leading nonrelativistic 
approximation.
Again, following the notation introduced in~\cite{JePa1996,JeSoMo1997},
the resulting additional contribution is
\begin{equation}
\label{E4bspec}
E_{4b} = \left(\frac{\alpha}{\pi}\right)^2 \, (Z\alpha)^4 \,
\frac{m}{n^3} \, \Delta_{\mathrm{fs}} F_{\delta y}\,.
\end{equation}
The sum of (\ref{E4aspec}) and (\ref{E4bspec}) is just the 
$(Z\alpha)^6$--component of the 
fine-structure difference of the one-loop self energy 
from~\cite{JePa1996,JeSoMo1997}, multiplied
by an additional factor $\alpha/\pi$. It can also be written as
\begin{equation}
\label{E4}
E_4 = E_{4a} + E_{4b} = \left( \frac{\alpha}{\pi} \right)^2 \,
\frac{(Z\alpha)^6 \, m}{n^3} \, \left[ - \frac{n^2 - 1}{3 n^2} \, 
  \ln \frac{2 \epsilon}{(Z\alpha)^2\,m} 
     + \frac{n^2 - 1}{n^2} \, \Delta_{\mathrm{fs}}{{\ell}}_4(n) \right]\,,
\end{equation}
where $\Delta_{\mathrm{fs}}{{\ell}}_4(n)$ could be interpreted as a 
relativistic generalization of a
Bethe logarithm, which is $n$-dependent. However, a significant
numerical fraction of the $n$-dependence can be eliminated 
if the factor $(n^2 - 1)/n^2$ is taken out of the 
final result. The evaluation of $\Delta_{\mathrm{fs}}{{\ell}}_4(n)$ 
has recently been performed 
in~\cite{JeLBMoSoIn2001} with improved numerical methods 
(see e.g.~\cite{JeMoSoWe1999}), and the following results have been obtained:
\begin{eqnarray}
\Delta_{\mathrm{fs}}{\ell}_4(2) &=& 0.512~559~768(1)\,, \nonumber\\
\Delta_{\mathrm{fs}}{{\ell}}_4(3) &=& 0.511~978~815(1)\,, \nonumber\\
\Delta_{\mathrm{fs}}{{\ell}}_4(4) &=& 0.516~095~539(1)\,, \nonumber\\
\Delta_{\mathrm{fs}}{{\ell}}_4(5) &=& 0.519~976~941(1)\,, 
\end{eqnarray}
where the uncertainty is due to numerical integration.

There is, as stated above, a further
correction due to the explicit modification of the 
transition current due to the anomalous magnetic moment;
it can be obtained through the replacement
\begin{equation}
\alpha^i \to \alpha^i + F_2(0) \, 
\frac{{\mathrm i}\,\beta\,\sigma^{i\nu}}{2 m} \, q_\nu
\end{equation}
and must be considered {\em in addition} to the correction (\ref{FWcorrection}).
A careful consideration of the nonrelativistic limit of this 
correction to the current, including retardation effects, leads to the
result
\begin{equation}
\delta \bbox{j}_{5} = \frac{\alpha}{2 \pi} \, \frac{Z\alpha}{2 m r^3} \,\,
\bbox{\sigma} \times \bbox{r}\,.
\end{equation}
Consequently, we find that the correction is
effectively $F_2(0)$ times the retardation corrections to the
transition current $F_{\delta y}$ found in~\cite{JePa1996,JeSoMo1997}.
We obtain
\begin{equation}
E_5 = \left(\frac{\alpha}{\pi}\right)^2 \, (Z\alpha)^4 \,
\frac{m}{n^3} \, \frac{\Delta_{\mathrm{fs}} F_{\delta y}}{2}\,.
\end{equation}
In analogy with $E_4$, this correction can favorably be rewritten as
\begin{equation}
\label{E5}
E_5 = \left( \frac{\alpha}{\pi} \right)^2 \,
\frac{(Z\alpha)^6 \, m}{n^3} \, \left[ 
\frac{n^2 - 1}{n^2} \, \Delta_{\mathrm{fs}}{{\ell}}_5(n) \right]\,,
\end{equation}
On the basis of~\cite{JePa1996,JeSoMo1997,JeLBMoSoIn2001}, we obtain
\begin{eqnarray}
\Delta_{\mathrm{fs}}{{\ell}}_5(2) &=& -0.173~344~868(1)\,, \nonumber\\
\Delta_{\mathrm{fs}}{{\ell}}_5(3) &=& -0.164~776~514(1)\,, \nonumber\\
\Delta_{\mathrm{fs}}{{\ell}}_5(4) &=& -0.162~263~216(1)\,, \nonumber\\
\Delta_{\mathrm{fs}}{{\ell}}_5(5) &=& -0.161~165~602(1)\,. 
\end{eqnarray}
The final result for the low-energy part is
\begin{equation}
\label{EL}
E_{\mathrm L} = E_4 + E_5\,,
\end{equation}
with $E_4$ and $E_5$ being given in eqs.~(\ref{E4}) and (\ref{E5}),
respectively. 

We can now understand why it was possible to join the two contributions
with ``mixed'' and ``low-and-low'' energy virtual photons (ii) and (iii),
which were discussed in sec.~\ref{Introduction}, into a joint 
``low-energy part''. The reason is simple:
The effective hamiltonian (\ref{Heff}) has no
infrared divergence, because it involves the low-energy limit
of the magnetic form factor $F_2$, which is infrared safe in
one-loop order according to eq.~(\ref{yefrsu}). Because the main
contribution to the quantity $F_2(0)$ is caused by hard virtual photons,
it is also justified to say that the contribution of ``low-and-low''
energy virtual photons vanishes at the order of interest for the 
current calculation (fine-structure difference). In higher-loop order,
the further infrared divergence acquired by $F_2$ would lead to an infrared
divergence in the effective hamiltonian constructed in analogy with 
eq.~(\ref{Heff}); this infrared divergence would have 
to be attributed to a ``mixed'' contribution (one photon of high
energy, and one low-energy photon). 

%
%
\section{Results and Conclusions}
\label{resconcl}

We have obtained analytic results for higher-order correction to the two-loop 
self-energy of P states in hydrogen-like systems.  
In our calculation,
we have analyzed the electron form factors through two-loop
order in sec.~\ref{twoloopff}, and we have split the calculation into 
a high-energy part with two hard virtual photons discussed in
sec.~\ref{hep}, and a low-energy part with at least one soft virtual
photon analyzed in sec.~\ref{lep}.
The final result for the contribution to the 
fine-structure energy difference 
is obtained by adding the high-energy contributions 
$E_1$ -- $E_3$ given in
eqs.~(\ref{E1}), (\ref{E2a}), (\ref{E2b}), (\ref{E3}), 
and the low-energy effects 
$E_4$ and $E_5$ from eqs.~(\ref{E4}) and (\ref{E5}).
The dependence on $\epsilon$ cancels out in the final result
which is the sum of the high-energy part $E_{\mathrm H}$
given in eq.~(\ref{EH}) and the low-energy part 
$E_{\mathrm L}$ defined in eq.~(\ref{EL}).
This is also evident when considering explicitly
the eqs.~(\ref{E2b}) and (\ref{E4}). The final results for the analytic
coefficients of order $\alpha^2 (Z\alpha)^6$ read 
\begin{equation}
\Delta_{\mathrm{fs}} B_{\mathrm 61} =
  - \frac{n^2 - 1}{3 n^2} \,.
\end{equation}
[see also eq.~(\ref{DeltaB61})] and
\begin{eqnarray}
\label{final}
\lefteqn{\Delta_{\mathrm{fs}} B_{\mathrm 60} =
   \left( \frac{227}{2880} + \frac{1}{32 n} - \frac{3}{80 n^2} \right) +
   F_2^{(4),S}(0) \, \left( \frac{487}{720} + \frac{5}{8 n} -
        \frac{23}{20 n^2} \right)}
\nonumber\\[1ex]
& &  + \frac{n^2 - 1}{n^2} \,
       \left[- \left( F_1'^{(4),S}(0) + 
            2\, {\mathcal F}_2'^{(4),S}(0) \right) \, m^2 
          - \frac{5}{18} + \Delta_{\mathrm{fs}}{{\ell}}_4(n) +
            \Delta_{\mathrm{fs}}{{\ell}}_5(n) \right]\,,
\end{eqnarray}
where explicit numerical results for $F_1'^{(4),S}(0)$,
$F_2^{(4),S}(0)$ and ${\mathcal F}_2'^{(4),S}(0)$ can be found in 
eqs.~(\ref{seF1p0}), (\ref{seF20}) and (\ref{seF2p0}),
respectively. This result refers to the pure self-energy diagrams
in fig.~\ref{fig1}. The result reads numerically for the principal
quantum numbers $n=2$--5,
\begin{eqnarray}
\Delta_{\mathrm{fs}} B_{\mathrm 60}(2) &=& -0.361~196~470(1)\,,\\
\Delta_{\mathrm{fs}} B_{\mathrm 60}(3) &=& -0.411~156~068(1)\,,\\
\Delta_{\mathrm{fs}} B_{\mathrm 60}(4) &=& -0.419~926~624(1)\,,\\
\Delta_{\mathrm{fs}} B_{\mathrm 60}(5) &=& -0.419~832~876(1)\,.
\end{eqnarray}
If it is desired to add in the combined self-energy
vacuum-polarization diagram from fig.~\ref{fig2}, then the 
form-factor results from eqs.~(\ref{resultF1prime}), (\ref{resultF2}) and
(\ref{resultF2prime}) instead of
the pure self-energy results given in eqs.~(\ref{seF1p0}), (\ref{seF20}) 
and (\ref{seF2p0}) have to be used in evaluating (\ref{final}).
When including the combined self-energy
vacuum-polarization diagram from fig.~\ref{fig2},
there is no further low-energy contribution, so that
the alternative set of numerical values for the form factors
from eqs.~(\ref{resultF1prime}), (\ref{resultF2}) and
(\ref{resultF2prime}) fully takes into account the additional 
effect of the diagram in fig.~\ref{fig1} on the fine-structure 
in the order of $\alpha^2\,(Z\alpha)^6$.

It is perhaps worth mentioning that 
for the one-loop self-energy, analytic
coefficients are known only up to the order of 
$\alpha\,(Z\alpha)^6$~\cite{JePa1996},
but the remaining uncertainty is removed by recent non-perturbative
numerical calculations~\cite{JeMoSo1999,JeMoSo2001pra,JeLBMoSoIn2001}.   
For the two-loop effect, the $(Z\alpha)$-expansion 
converges more rapidly than for the 
one-loop effect in absolute frequency units because of the 
additional radiative factor $\alpha/\pi$ which decreases the overall size 
of the effect.

It is hoped that the analytic calculations for low nuclear
charge number $Z$ will be supplemented 
in the future by an accurate numerical treatment of the two-loop
self-energy problem (see also related recent work in the 
high-$Z$ region, refs.~\cite{Ye2000,ShYe2001,GoEtAl2001}). This presupposes 
that the considerable numerical 
problems in the domain of small nuclear charge
could be solved by adequate numerical methods, and that the further
problem of the increased computational demand of the two-loop effect
in comparison to the one-loop problem~\cite{JeMoSo1999,JeMoSo2001pra}
can be tackled -- possibly by massively parallel computer architectures.
Note, however, that the most accurate theoretical predictions could
only be reached in combining numerical and analytic results.
The reason is the following:
All numerical calculations are performed in the non-recoil limit
which is the limit of infinite nuclear mass. This is not quite
sufficient for an accurate theoretical treatment because
the self-energy effect for a bound-state depends genuinely on the ratio of 
the orbiting particle to the nuclear mass -- an effect beyond the 
recoil correction. For example, the argument of the logarithms
in (\ref{DefESE}) should be replaced according to
$\ln[(Z\alpha)^{-2}] \to \ln[\sigma\,(Z\alpha)^{-2}]$, 
where $\sigma = m/m_{\mathrm r}$ and $m_{\mathrm r}$ is the reduced
mass~\cite{SaYe1990}. The possibility to include these 
tiny, but important effects 
depends crucially on a reliable knowledge of the analytic 
coefficients {\em in combination} 
with an accurate numerical treatment of the problem.

The analytic results can be used
to obtain improved theoretical predictions 
for the hydrogenic fine structure
as compared to the previous 
order--$\alpha^7$--calculations~\cite{JePa1996,JeSoMo1997},
because they remove the principal theoretical uncertainty 
in the order of $\alpha^8$
due to the problematic two-loop self-energy which is 
represented diagrammatically in fig.~\ref{fig1}.
A compilation of the other corrections relevant at the order of 
$\alpha^8$, including but not limited to the vacuum polarization effects, 
whose evaluation is rather straightforward, 
will be presented elsewhere. 
Our calculation illustrates the usefulness of the simplified
effective treatments of two-loop effects in the analytic approach
based on the modified Dirac hamiltonian (\ref{HDm}) and the 
``$\epsilon$ method'' (see~\cite{Pa1993,JePa1996,JeSoMo1997,Pa1998} 
and~app.~\ref{appa}). This aspect highlights, as we believe, the 
need for systematic, simplified treatments of higher-order radiative
corrections in bound systems.  

In this paper, we primarily address spin-dependent effects in
one-electron (hydrogenlike) systems. However, the same effects also
contribute to the fine-structure splitting in two-electron (heliumlike)
systems. There is currently remarkable interest in improved measurements
of the fine-structure splitting in helium and heliumlike atomic systems
with low nuclear charge~\cite{heliumlike}. The effects addressed in this
paper contribute to the fine-structure splitting in helium on the level of
100~Hz, which is not much smaller than the current experimental accuracy
of about 1~kHz, and allows for an estimate of uncalculated yet
higher-order contributions.

The results of this paper are a step in a systematic study of the
higher-order binding corrections to the two-loop Lamb shift of S and P
states. The scheme of calculation permits not the only a simplified
treatment of the problem via a separation into appropriate energy regions
for the two virtual photons, but also a clear identification of the
spin-independent and the spin-dependent contributions to the self-energy.
The results will therefore be directly applicable to the total coefficients
for P states once the spin-independent parts (the two-loop Bethe
logarithms) are calculated. These are currently under study.

%
%
\section*{Acknowledgments}

U.D.J. acknowledges helpful conversations with Professor Gerhard
Soff. This work was supported in part by a research grant
from the Polish Committee for Scientific Research under
contract no. 2P03B 057 18.

\appendix

%
%
\section{The ``$\epsilon$ Method''}
\label{appa}

We discuss here, by way of example,
the $\epsilon$ method employed
in the analytic calculation of self-energy effects in bound systems.
This method is very suitable~\cite{JePa1996,JeSoMo1997}
for the separation of the two different
energy scales for virtual photons: the nonrelativistic domain, in which the 
virtual photon assumes values of the order of the atomic binding energy,
and the relativistic domain, in which the virtual photon assumes values
of the order of the electron rest mass. Different approximation schemes and
different asymptotic expansions are
adequate for the two different domains. Without these approximations and
expansions, the analytic evaluation of either the high- or the low-energy
part would not be feasible. 
At the same time, the model example discussed in this appendix
is meant to illustrate the usefulness of the ``$\epsilon$ method''
in a more general context.

We will consider here a model problem with only one ``virtual photon''.
The separation into high-and low-energy photons necessitates the temporary
introduction of a parameter $\epsilon$; the dependence on $\epsilon$ cancels  
when the high- and the low-energy parts are added together. We have,
\begin{eqnarray}
\mbox{nonrelativistic domain} & \ll \epsilon \ll & 
\mbox{electron rest mass} \\
(Z \alpha)^2 \, m_{\mathrm{e}} & \ll \epsilon \ll & m_{\mathrm{e}}\,,
\end{eqnarray}
where $\alpha$ is the fine structure constant, 
and $Z$ is the nuclear charge. The high-energy part
is associated with photon energies $\omega > \epsilon$, and the 
low-energy part is associated with photon energies $\omega < \epsilon$.

In order to illustrate the procedure, we discuss a simple,
one-dimensional example: the evaluation of 
\begin{equation}
I(\beta) = 
\int_0^1 \sqrt{\frac{\omega^2 + \beta^2}{1 - \omega^2}} \, 
{\mathrm{d}}\omega\,.
\end{equation}
where the integration variable $\omega$ might be interpreted as the 
``energy'' of a ``virtual photon''. The integral $I$ can be expressed in
terms of special functions,
\begin{equation}
I(\beta) = \beta\,E\left(-\frac{1}{\beta^2}\right) = 
\beta \, \frac{\pi}{2} \,
{}_2 F_1 \left(-\frac{1}{2}, \frac{1}{2}; 1;  -\frac{1}{\beta^2}\right)\,,
\end{equation}
where $E$ is the complete elliptic integral of the second kind, and 
${}_2 F_1$ denotes a hypergeometric function. An alternative integral
representation reads $I(\beta) = \int_0^{\pi/2} \sqrt{\beta^2 + 
\sin^2(\omega)} \, {\mathrm{d}}\omega$.

The purpose of the calculation is
to derive a semi-analytic expansion of $I(\beta)$ in powers of 
$\beta$ and $\ln \beta$. The fine structure constant $\alpha$ takes the 
r\^{o}le 
of the expansion parameter $\beta$ in actual self-energy calculations. 
We discuss first the ``high-energy part'' of the calculation. 
It is given by the expression
\begin{eqnarray}
\label{IH}
I_{\mathrm{H}}(\beta) &=& 
\int_\epsilon^1 \sqrt{\frac{\omega^2 + \beta^2}{1 - \omega^2}} \,
{\mathrm{d}}\omega\,.
\end{eqnarray}
For $\omega > \epsilon$, we may expand
\begin{equation}
\label{expansion1}
\sqrt{\omega^2 + \beta^2} = \omega + \frac{\beta^2}{2\,\omega} +
  \frac{\beta^4}{8\,\omega^3} + \mathcal{O}(\beta^6)\,,
\end{equation}
but this expansion is not applicable in higher
orders to the domain $0 < \omega < \epsilon$ because of the appearance
of inverse powers of $\omega$ (analogous to an
``infrared divergence'' in QED).

The separation parameter $\epsilon$ acts an infrared regulator.
After expanding in $\beta$ [see Eq.~(\ref{expansion1})], the 
resulting integrals in each order of $\beta$ can be evaluated
analytically. Subsequently, we expand every term in
the $\beta$-expansion in powers of $\epsilon$ up to
the order $\epsilon^0$, i.e.~we keep only the divergent and constant
terms in $\epsilon$.
The result is
\begin{eqnarray}
\label{IHresult}
I_{\mathrm{H}}(\beta,\epsilon) &=& 1 + \beta^2 \, \left\{ \frac{1}{2} \,
\ln \left(\frac{2}{\epsilon}\right) +
{\mathcal{O}}(\epsilon) \right\} \nonumber\\[1ex]
& & + \beta^4 \, \left\{ - \frac{1}{16\, \epsilon^2} -
\frac{1}{16} \, \ln\left( \frac{2}{\epsilon} \right) +
\frac{1}{32} + {\mathcal{O}}(\epsilon) \right\} 
\nonumber\\[1ex]
& & + \beta^6 \, \left\{ \frac{1}{64\, \epsilon^4} +
\frac{1}{64 \, \epsilon^2} +
\frac{3}{128} \, \ln\left( \frac{2}{\epsilon} \right) - \frac{7}{512} +
{\mathcal{O}}(\epsilon) \right\} \nonumber\\[1ex]
& & + {\mathcal{O}}(\beta^8) \,.
\end{eqnarray}
Here, the ``${\mathcal{O}}$''-symbol identifies a contribution for which
${\mathcal{O}}(x)/x \to {\mathrm{const.}}$ as 
$x \to 0$, whereas the ``${\mathrm{o}}$''-symbol identifies
the weaker requirement ${\mathrm{o}}(x) \to 0$ as 
$x \to 0$; this is consistent with the 
standard notation (see e.g.~\cite{Er1987}).

The contribution $I_{\mathrm{H}}(\beta)$ corresponds to the 
``high-energy part'' in analytic self-energy calculations, where
the propagator of the bound electron is explicitly expanded in
powers of the fine structure constant $\alpha$.
Now we turn to the ``low-energy part''. The expression for the 
low-energy part ($0 < \omega < \epsilon$) reads
\begin{eqnarray}
\label{IL}
I_{\mathrm{L}}(\beta) &=& 
\int_0^\epsilon \sqrt{\frac{\omega^2 + \beta^2}{1 - \omega^2}} \,
{\mathrm{d}}\omega\,.
\end{eqnarray}
The expansion (\ref{expansion1}) is not applicable in this energy
domain; we therefore have to keep the numerator of the integrand
$\sqrt{\omega^2 + \beta^2}$ in unexpanded form. However, we can expand
the denominator $\sqrt{1 - \omega^2}$ of the integrand in powers
of $\omega$; because $0 < \omega <\epsilon$ (with $\epsilon$ small),
this expansion in $\omega$ is in fact an expansion in $\beta$ --
although the situation is somewhat problematic in the sense that
every term in the $\omega$-expansion gives rise to terms of arbitrarily
high order in the $\beta$-expansion [see also Eq.~(\ref{problematic})
below]. 

The term $\sqrt{\omega^2 + \beta^2}$ is analogous to the 
Schr\"{o}dinger--Coulomb propagator in the self-energy calculation
which has to be kept in unexpanded form, whereas the expansion
\begin{equation}
\label{expansion2}
\frac{1}{\sqrt{1 - \omega^2}} = 1 + \frac{\omega^2}{2} + 
\frac{3}{8}\,\omega^4 + {\mathcal{O}}(\omega^6)
\end{equation}
corresponds to the expansion into the $(Z\alpha)$-expansion
in the low-energy part.

Every term in the expansion (\ref{expansion2}) gives rise to arbitrarily
high-order corrections in $\beta$, but it starts with the power
$\omega^n \to \beta^{n+2}$. For example, we have for the leading
term of order $\omega^0 = 1$ from Eq.~(\ref{expansion2}),
\begin{eqnarray}
\label{problematic}
\lefteqn{\int_0^\epsilon \sqrt{\omega^2 + \beta^2} \, {\mathrm{d}}\omega =
\beta^2 \, \left\{ \frac{1}{2} \ln\left(\frac{2}{\beta} \, \epsilon \right) + 
\frac{1}{4} + {\mathcal{O}}(\epsilon) \right\}} \nonumber\\[1ex]
& & + \beta^4 \, \left\{ \frac{1}{16 \, \epsilon^2} +
{\mathcal{O}}(\epsilon) \right\} +
\beta^6 \, \left\{ - \frac{1}{64 \, \epsilon^4} +
{\mathcal{O}}(\epsilon) \right\} + {\mathcal{O}}(\beta^8)\,.
\end{eqnarray}
Note that the terms generated in the orders $\beta^4$ and $\beta^6$
are needed to cancel divergent contributions in respective
orders of $\beta$ from the high-energy part given in
Eq.~(\ref{expansion1}). 
The term of order $\omega^2$ from (\ref{expansion2}) results in
\begin{eqnarray}
\frac{1}{2}\, \int_0^\epsilon \omega^2 \, \sqrt{\omega^2 + \beta^2} \, 
{\mathrm{d}}\omega &=&
\beta^4 \, \left\{ -\frac{1}{16} \, 
\ln\left(\frac{2}{\beta} \, \epsilon \right) + 
\frac{1}{64} + {\mathcal{O}}(\epsilon) \right\} \nonumber\\[1ex]
& & + \beta^6 \, \left\{ - \frac{1}{64 \, \epsilon^2} + 
{\mathcal{O}}(\epsilon) \right\} +
{\mathcal{O}}(\beta^8)\,.
\end{eqnarray}
Altogether, we obtain for the low-energy part,
\begin{eqnarray}
\label{ILresult}
\lefteqn{I_{\mathrm{L}}(\beta,\epsilon) = \beta^2 \, 
\left\{ \frac{1}{2} \, \ln \left(\frac{2}{\beta} \, \epsilon\right) +
\frac{1}{4} + {\mathcal{O}}(\epsilon) \right\}} \nonumber\\[1ex]
& & + \beta^4 \, \left\{ \frac{1}{16\, \epsilon^2} -
\frac{1}{16} \, \ln\left( \frac{2}{\beta} \, \epsilon \right) +
\frac{1}{64} + {\mathcal{O}}(\epsilon) \right\} 
\nonumber\\[1ex]
& & + \beta^6 \, \left\{ - \frac{1}{64\, \epsilon^4} -
\frac{1}{64 \, \epsilon^2} +
\frac{3}{128} \, \ln\left( \frac{2}{\beta} \, \epsilon \right) -
\frac{5}{512} +
{\mathcal{O}}(\epsilon) \right\} \nonumber\\[1ex]
& & + {\mathcal{O}}(\beta^8 \, \ln \beta) \,.
\end{eqnarray}
When the high-energy part (\ref{IHresult}) and the low-energy part 
(\ref{ILresult}) are added, the dependence on $\epsilon$ cancels,
and we have
\begin{eqnarray}
I(\beta) &=& I_{\mathrm{H}}(\beta, \epsilon) + 
I_{\mathrm{L}}(\beta, \epsilon) \nonumber\\[2ex]
&=& 1 + \beta^2 \,
\left\{ \frac{1}{2}\,
\ln\left(\frac{4}{\beta}\right) + \frac{1}{4} \right\}
\nonumber\\[2ex]
& & + \beta^4 \, \left\{ -\frac{1}{16}\, 
\ln\left(\frac{4}{\beta}\right) + \frac{3}{64} \right\}
\nonumber\\[2ex]
& & + \beta^6 \, \left\{ \frac{3}{128}\, 
\ln\left(\frac{4}{\beta}\right) - \frac{3}{128} \right\}
+ {\mathcal{O}}(\beta^8 \, \ln \beta)\,.
\end{eqnarray}
In order to illustrate the analogy with the self-energy calculation
presented here, we would like to point out that
the dependence on $\epsilon$ cancels out in the final result
which is the sum of the high-energy part $E_{\mathrm H}$
given in eq.~(\ref{EH}) and the low-energy part
$E_{\mathrm L}$ in eq.~(\ref{EL}).


\newpage

\begin{thebibliography}{10}

\bibitem{NiEtAl2000}
M. Niering, R. Holzwarth, J. Reichert, P. Pokasov, T. Udem, M. Weitz, T.~W.
  H\"{a}nsch, P. Lemonde, G. Santarelli, M. Abgrall, P. Laurent, C. Salomon,
  and A. Clairon, Phys. Rev. Lett. {\bf 84},  5496  (2000).

\bibitem{HaPr2001}
T.~W. H\"{a}nsch, private communication (2001).

\bibitem{SaYe1990}
J. Sapirstein and D.~R. Yennie,  in {\em Quantum Electrodynamics}, edited by T.
  Kinoshita (World Scientific, Singapore, 1990), pp.\ 560--672.

\bibitem{MoPlSo1998}
P.~J. Mohr, G. Plunien, and G. Soff, Phys. Rep. {\bf 293},  227  (1998).

\bibitem{ErYe1965ab}
G.~W. Erickson and D.~R. Yennie, Ann. Phys. (N. Y.) {\bf 35},  271, 447
  (1965).

\bibitem{Er1971}
G.~W. Erickson, Phys. Rev. Lett. {\bf 27},  780  (1971).

\bibitem{Sa1981}
J. Sapirstein, Phys. Rev. Lett. {\bf 47},  1723  (1981).

\bibitem{Pa1993}
K. Pachucki, Ann. Phys. (N. Y.) {\bf 226},  1  (1993).

\bibitem{JeMoSo1999}
U.~D. Jentschura, P.~J. Mohr, and G. Soff, Phys. Rev. Lett. {\bf 82},  53
  (1999).

\bibitem{Wo1988}
S. Wolfram, {\em Mathematica-A System for Doing Mathematics by Computer}
  (Addison-Wesley, Reading, MA, 1988).

\bibitem{disclaimer}
Certain commercial equipment, instruments, or materials are identified in this
  paper to foster understanding. Such identification does not imply
  recommendation or endorsement by the National Institute of Standards and
  Technology, nor does it imply that the materials or equipment identified are
  necessarily the best available for the purpose.

\bibitem{JePa1996}
U.~D. Jentschura and K. Pachucki, Phys. Rev. A {\bf 54},  1853  (1996).

\bibitem{JeSoMo1997}
U.~D. Jentschura, G. Soff, and P.~J. Mohr, Phys. Rev. A {\bf 56},  1739
  (1997).

\bibitem{Pa2001}
K. Pachucki, Phys. Rev. A {\bf 63},  042503  (2001).

\bibitem{ApBr1970}
E. Appelquist and S.~J. Brodsky, Phys. Rev. A {\bf 2},  2293  (1970).

\bibitem{Pa1994prl}
K. Pachucki, Phys. Rev. Lett. {\bf 72},  3154  (1994).

\bibitem{EiKaSh1995}
M. Eides and V. Shelyuto, Phys. Rev. A {\bf 52},  954  (1995).

\bibitem{Pa1999}
K. Pachucki, J. Phys. B {\bf 32},  137  (1999).

\bibitem{ItZu1980}
C. Itzykson and J.~B. Zuber, {\em Quantum Field Theory} (McGraw-Hill, New York,
  NY, 1980).

\bibitem{Pa1998}
K. Pachucki, J. Phys. B {\bf 31},  5123  (1998).

\bibitem{BB1984}
I. Bia\l{}ynicki-Birula,  in {\em Quantum Electrodynamics and Quantum Optics},
  edited by A.~O. Barut (Plenum, New York, 1984), pp.\ 41--61.

\bibitem{CaLe1986}
W.~E. Caswell and G.~P. Lepage, Phys. Lett. B {\bf 167},  437  (1986).

\bibitem{Re1972a}
E. Remiddi, Nuovo Cim. A {\bf 11},  825  (1972).

\bibitem{Re1972b}
E. Remiddi, Nuovo Cim. A {\bf 11},  865  (1972).

\bibitem{YeFrSu1961}
D.~R. Yennie, S.~C. Frautschi, and H. Suura, Ann. Phys. (N. Y.) {\bf 13},  379
  (1961).

\bibitem{MaRe2001} P.~Mastrolia and E.~Remiddi,
{\em private communication} (2001).

\bibitem{ReVa2000} E.~Remiddi and J.~A.~M.~Vermaseren, 
  Int.\ J.\ Mod.\ Phys.\ {\bf A15}, 725 (2000).

\bibitem{GeRe2001} T.~Gehrmann and E.~Remiddi,
  Comp.\ Phys.\ Commun.\ {\bf 141}, 296 (2001).

\bibitem{SwDr1991a}
R.~A. Swainson and G.~W.~F. Drake, J. Phys. A {\bf 24},  79  (1991).

\bibitem{SwDr1991b}
R.~A. Swainson and G.~W.~F. Drake, J. Phys. A {\bf 24},  95  (1991).

\bibitem{BeSa1957}
H.~A. Bethe and E.~E. Salpeter, {\em Quantum Mechanics of One- and Two-Electron
  Atoms} (Springer, Berlin, 1957).

\bibitem{JeLBMoSoIn2001}
U.~D. Jentschura, E.~O. LeBigot, P.~J. Mohr, G. Soff, and P.~J. Indelicato, in
  preparation, 2001.

\bibitem{JeMoSoWe1999}
U.~D. Jentschura, P.~J. Mohr, G. Soff, and E.~J. Weniger, Comput. Phys. Commun.
  {\bf 116},  28  (1999).

\bibitem{JeMoSo2001pra}
U.~D. Jentschura, P.~J. Mohr, and G. Soff, Phys. Rev. A {\bf 64},  042512
  (2001).

\bibitem{Ye2000}
V.~A. Yerokhin, Phys. Rev. A {\bf 62},  012508  (2000).

\bibitem{ShYe2001}
V.~A. Yerokhin and V.~M. Shabaev, Phys. Rev. A {\bf 64},
  062507 (2001); e-print hep-ph/0107036.

\bibitem{GoEtAl2001}
I.~A. Goidenko, L.~N. Labzowsky, A.~V. Nefiodov, U.~D. Jentschura, G. Plunien,
  S. Zschocke, and G. Soff, Phys. Scr. T {\bf 92},  426  (2001).

\bibitem{Er1987}
H. Erdelyi, {\em Asymptotic Expansions} (Dover, New York, NY, 1987).

\bibitem{heliumlike} E. G. Myers and M. R. Tabutt,
  Phys. Rev. A {\bf 61}, 010501 (1999);
E. G. Myers, H. S. Margolis, J. K. Thompson, M. A. Farmer,
  J. D. Silver, and M. R. Tabutt,
  Phys. Rev. Lett. {\bf 82}, 4200 (1999);
C. H. Storry, M. C. George, and E. A. Hessels,
  Phys. Rev. Lett. {\bf 84}, 3274 (2000);
M. C. George, L. D. Lombardi, and E. A. Hessels,
  Phys. Rev. Lett. {\bf 87}, 173002 (2001). 
 
\end{thebibliography}
\end{document}